\documentclass[a4paper,oneside,12pt,DIV=11,parskip=half]{scrartcl}
\usepackage[T1]{fontenc}
\usepackage{tgschola}
\usepackage{fouriernc}
\usepackage{amsmath}
\usepackage{hyperref}
\usepackage{makeidx}
\usepackage{axodraw2}

\addtokomafont{disposition}{\normalfont\scshape}
\addtokomafont{section}{\centering\normalsize\bfseries\boldmath\MakeUppercase}
\addtokomafont{title}{\normalfont\scshape}

\allowdisplaybreaks

\makeindex

\hypersetup{colorlinks,citecolor=ForestGreen}

\usepackage{xfrac}
\usepackage{xspace}

\newcommand{\eps}{\epsilon}
\newcommand{\D}{\textrm{d}}
\newcommand{\PS}{\textrm{PS}}
\newcommand{\nn}{\nonumber}
\newcommand{\x}{\,}

\newcommand{\fire}{\texttt{FIRE5}\xspace}
\newcommand{\litered}{\texttt{LiteRed}\xspace}
\newcommand{\dream}{\texttt{DREAM}\xspace}
\newcommand{\summertime}{\texttt{SummerTime}\xspace}

\def\z(#1){
 \zeta_{#1}
}

\def\num(#1){
  \begin{axopicture}(100,55)(0,-2)
  \textcolor[rgb]{0.4,0.4,0.4}{\PText(86,0)(0)[c]{\footnotesize $$}}
}
\def\dot(#1,#2){
  \White{\Vertex(#1,#2){3}}
}
\def\den(#1,#2,#3,#4){
  \Line[color=BrickRed,width=0.85](#1,#2)(#3,#4)
  \Vertex(#1,#2){1.5}
  \Vertex(#3,#4){1.5}
}
\def\cut(#1,#2,#3,#4){
  \Line[color=Gray,width=0.5,dash](#1,#2)(#3,#4)
  \Vertex(#1,#2){1.5}
  \Vertex(#3,#4){1.5}
}
\def\cutd(#1,#2,#3,#4){
  \DoubleLine[color=Gray,width=0.5,dash](#1,#2)(#3,#4){1.5}
  \Vertex(#1,#2){1.5}
  \Vertex(#3,#4){1.5}
}
\def\arc(#1,#2,#3,#4,#5){
  \Arc[color=Gray,width=0.5,dash](#1,#2)(#3,#4,#5)
}

\hypersetup{
    pdftitle = {Five-Particle Phase-Space Integrals in QCD},
    pdfauthor = {O. Gituliar, V. Magerya, A. Pikelner}
}

\begin{document}

\let\endtitlepage\relax
\begin{titlepage}
\hfill DESY 18-043
  \mbox{}\vspace{1.5cm}
  \begin{center}
    {\LARGE Five-Particle Phase-Space Integrals in QCD}\par
      \vspace{2mm}
    {{\scshape O.~Gituliar} \\ \href{mailto:oleksandr@gituliar.net}{oleksandr@gituliar.net}}\par
    {{\scshape V.~Magerya} \\ \href{mailto:vitaly.magerya@tx97.net}{vitaly.magerya@tx97.net}}\par
    {{\scshape A.~Pikelner\footnote{On leave of absence from Joint Institute for Nuclear
    Research, 141980 Dubna, Russia.}} \\ \href{mailto:andrey.pikelner@desy.de}{andrey.pikelner@desy.de}}\par
      \vspace{2mm}
    {\small \em II. Institut f\"ur Theoretische Physik, Universit\"at Hamburg,\\ Luruper Chaussee 149, D-22761 Hamburg, Germany}\par
    {\small\today}\par
    \vspace{5mm}
  \end{center}
\end{titlepage}

\begin{abstract}
We present analytical expressions for the 31 five-particle phase-space master
integrals in massless QCD as an $\eps$-series with coefficients being multiple
zeta values of weight up to 12. In addition, we provide computer code for the
Monte-Carlo integration in higher dimensions, based on the RAMBO algorithm,
that has been used to numerically cross-check the obtained results in 4, 6, and
8 dimensions.
\end{abstract}

\section{Introduction}

Nowadays, perturbative calculations play the key role in describing data from
high-energy particle colliders, such as the LHC, as well as in improving the
precision of numerical parameters in Standard Model and other models. It is
clear now that higher-order calculations will play an even more crucial role in
processing data from future high-luminosity colliders, like the FCC or the ILC,
where theoretical errors will dominate over experimental statistical errors.
These arguments motivate us to make one step forward beyond available fully-
inclusive phase-space integrals for a four-particle decay~\cite{GGH03} and
calculate a set of yet unknown integrals that corresponds to a five-particle
decay of a color-neutral off-shell particle in Quantum Chromodynamics. These
results, among others, are necessary ingredients in calculating various
exclusive quantities with the method of differential equations where they
are needed to determine integration constants, as for example discussed in
\cite{Git15, GM15} for the three-loop time-like splitting functions \cite{AMV11}.

In this paper, we focus on the calculation of master integrals that can be used
to express any other integral of the corresponding topology provided a set of
integration-by-parts rules (IBP) \cite{CT81} is known. Our approach is based
on techniques for solving dimensional recurrence relations (DRR) \cite{Tar96}
described in \cite{LM17a,LM17b}. In particular, we use \dream package
\cite{LM17a} to obtain numerical results for desired integrals with 2000-digit
precision and restore analytical form in terms of multiple zeta values (MZV)
\cite{Fur03,BBV09,LM15} up to weight 12 using the PSLQ method \cite{FBA99} as
implemented in Mathematica. We also present a Monte-Carlo code, based on the
RAMBO algorithm \cite{KSE85}, for numerical integration of the phase-space
integrals in arbitrary (integer) number of dimensions that has been used to
check consistency of the obtained results.


This paper is organized as following. In Section~\ref{sec:2} we introduce
our notation and describe our calculational method in more details. In
Section~\ref{sec:4} we provide complete results for four-particle integrals
and discuss numerical cross-checks using Monte-Carlo integration. In
Section~\ref{sec:5} we make final remarks. In Appendix~\ref{app:res} we provide
the complete list of master integrals.

Additionally we provide auxiliary files on arXiv\footnote{\url{https:// arxiv.org}}
containing the complete master integrals with MZV weight up to 12 as well as
the Monte-Carlo integration routines with the corresponding results.

\section{The Method}
\label{sec:2}

\begin{table}
\footnotesize
\begin{tabular}{c c c c}
\num(1)
\Line(10,20)(20,20)
\Line(80,20)(90,20)
\arc(50,10,31,20,160)
\arc(50,-15,46,50,130)
\arc(50,55,46,-130,-50)
\arc(50,30,31,-160,-20)
\cut(20,20,80,20)
\end{axopicture}
 & \num(2)
\Line(10,20)(20,20)
\Line(80,20)(90,20)
\den(20,20,40,40)
\cut(40,40,60,40)
\den(60,40,80,20)
\den(20,20,40,0)
\cut(40,0,60,0)
\den(60,0,80,20)
\cut(40,40,60,0)
\cut(60,40,40,0)
\cut(20,20,80,20)
\end{axopicture}
 & \num(3)
\Line(10,20)(20,20)
\Line(80,20)(90,20)
\den(20,20,50,40)
\cut(50,40,80,20)
\cut(20,20,50,0)
\den(50,0,80,20)

\arc(30,20,27,-45,45)
\arc(70,20,27,135,215)
\cut(50,40,50,0)
\end{axopicture}
 & \num(4)
\Line(10,20)(20,20)
\Line(80,20)(90,20)
\den(20,20,40,40)
\den(40,40,60,40)
\cut(60,40,80,20)
\cut(20,20,40,0)
\den(40,0,60,0)
\den(60,0,80,20)
\cutd(40,40,60,0)
\cut(60,40,40,0)
\end{axopicture}

\\ $F_{1}$ & $F_{2}$: 12 14 23 34 & $F_{3}$: 01 05 & $F_{4}$: 12 13 02 03
\\
\num(5)
\Line(10,20)(20,20)
\Line(80,20)(90,20)
\den(20,20,40,40)
\den(40,40,60,40)
\cut(60,40,80,20)
\cut(20,20,40,0)
\den(40,0,60,0)
\den(60,0,80,20)
\cut(40,40,40,0)
\cut(60,40,60,0)
\cut(40,40,60,0)
\end{axopicture}
 & \num(6)
\Line(10,20)(20,20)
\Line(80,20)(90,20)
\den(20,20,40,40)
\den(40,40,60,40)
\cut(60,40,80,20)
\cut(20,20,40,0)
\den(40,0,60,0)
\den(60,0,80,20)
\den(40,40,40,20)
\cut(40,20,40,0)
\cut(60,40,60,20)
\den(60,20,60,0)
\cut(40,20,60,20)
\end{axopicture}
 & \num(7)
\Line(10,20)(20,20)
\Line(80,20)(90,20)
\den(20,20,50,40)
\cut(50,40,80,20)
\cut(20,20,50,0)
\den(50,0,80,20)

\cut(20,20,80,20)
\dot(50,20)
\cutd(50,40,50,0)
\end{axopicture}
 & \num(8)
\Line(10,20)(20,20)
\Line(80,20)(90,20)
\den(20,20,40,40)
\cut(40,40,60,40)
\den(60,40,80,20)
\cut(20,20,40,0)
\cut(40,0,60,0)
\cut(60,0,80,20)
\den(40,40,60,0)
\cut(20,20,80,20)
\dot(50,20)
\den(60,40,40,0)
\end{axopicture}

\\ $F_{5}$: 12 34 02 03 & $F_{6}$: 12 13 02 04 45 35 & $F_{7}$: 123 124 & $F_{8}$: 13 14 123 124
\\
\num(9)
\Line(10,20)(20,20)
\Line(80,20)(90,20)
\den(20,20,40,40)
\cut(40,40,60,40)
\den(60,40,80,20)
\den(20,20,40,0)
\cut(40,0,60,0)
\den(60,0,80,20)
\cut(40,40,60,0)
\cutd(60,40,40,0)
\end{axopicture}
 & \num(10)
\Line(10,20)(20,20)
\Line(80,20)(90,20)
\cutd(20,20,50,40)
\den(50,40,80,20)
\den(20,20,50,0)
\cut(50,0,80,20)

\cutd(50,40,50,0)
\end{axopicture}
 & \num(11)
\Line(10,20)(20,20)
\Line(80,20)(90,20)
\den(20,20,40,40)
\den(40,40,60,40)
\cut(60,40,80,20)
\cut(20,20,40,0)
\den(40,0,60,0)
\den(60,0,80,20)
\cut(40,40,60,0)
\cutd(60,40,40,0)
\end{axopicture}
 & \num(12)
\Line(10,20)(20,20)
\Line(80,20)(90,20)
\den(20,20,50,40)
\cut(50,40,80,20)
\cut(20,20,40,0)
\den(40,0,60,0)
\den(60,0,80,20)
\cut(50,40,40,0)
\cut(50,40,60,0)
\cut(60,0,20,20)
\end{axopicture}

\\ $F_{9}$: 12 23 023 012 & $F_{10}$: 01 123 & $F_{11}$: 01 123 014 05 & $F_{12}$: 12 023 05
\\
\num(13)
\Line(10,20)(20,20)
\Line(80,20)(90,20)
\den(20,20,30,35)
\den(30,35,50,40)
\den(50,40,70,35)
\cut(70,35,80,20)
\den(70,5,80,20)
\den(50,0,70,5)
\cut(50,0,70,35)
\cut(30,35,50,0)
\cut(70,5,50,40)
\cut(20,20,70,5)
\end{axopicture}
 & \num(14)
\Line(10,20)(20,20)
\Line(80,20)(90,20)
\cut(20,20,40,40)
\cut(40,40,60,40)
\cut(60,40,80,20)
\den(20,20,40,0)
\cut(40,0,60,0)
\den(60,0,80,20)
\den(40,40,60,0)
\cut(40,0,40,40)
\dot(50,20)
\den(60,40,40,0)
\end{axopicture}
 & \num(15)
\Line(10,20)(20,20)
\Line(80,20)(90,20)
\den(20,20,30,35)
\den(30,35,50,40)
\den(50,40,70,35)
\cut(70,35,80,20)
\cut(20,20,30,5)
\den(30,5,50,0)
\den(50,0,70,5)
\den(70,5,80,20)
\cut(30,35,70,5)
\cut(30,5,70,35)
\cut(50,0,50,40)
\end{axopicture}
 & \num(16)
\Line(10,20)(20,20)
\Line(80,20)(90,20)
\den(20,20,30,35)
\den(30,35,50,40)
\den(50,40,70,35)
\cut(70,35,80,20)
\cut(20,20,30,5)
\den(30,5,50,0)
\den(50,0,70,5)
\den(70,5,80,20)
\cut(30,35,70,5)
\cut(30,5,50,40)
\cut(50,0,70,35)
\end{axopicture}

\\ $F_{13}$: 12 13 124 05 02 & $F_{14}$: 02 12 023 05 & $F_{15}$: 12 13 124 134 02 03 & $F_{16}$: 12 34 02 03 124 134
\\
\num(17)
\Line(10,20)(20,20)
\Line(80,20)(90,20)
\den(20,20,50,40)
\cut(50,40,80,20)
\den(20,20,50,0)
\cut(50,0,80,20)
\cut(50,40,50,20)
\cut(50,20,50,0)
\cut(20,20,50,20)
\den(50,20,80,20)
\end{axopicture}
 & \num(18)
\Line(10,20)(20,20)
\Line(80,20)(90,20)
\cut(20,20,30,35)
\den(30,35,50,40)
\cut(50,40,70,35)
\cut(70,35,80,20)
\den(20,20,30,5)
\den(30,5,50,0)
\cut(50,0,70,5)
\den(70,5,80,20)
\cut(50,0,50,40)
\den(30,35,70,5)
\dot(50,20)
\den(30,5,70,35)
\end{axopicture}
 & \num(19)
\Line(10,20)(20,20)
\Line(80,20)(90,20)
\cut(20,20,30,35)
\cut(30,35,50,40)
\cut(50,40,70,35)
\den(70,35,80,20)
\den(20,20,30,5)
\den(30,5,50,0)
\cut(50,0,70,5)
\den(70,5,80,20)
\den(30,35,70,5)
\cut(30,5,70,35)
\dot(50,20)
\den(50,0,50,40)
\end{axopicture}
 & \num(20)
\Line(10,20)(20,20)
\Line(80,20)(90,20)
\den(20,20,30,35)
\cut(30,35,50,40)
\den(50,40,70,35)
\den(70,35,80,20)
\den(20,20,30,5)
\den(30,5,50,0)
\cut(50,0,70,5)
\den(70,5,80,20)
\cut(30,35,70,5)
\cut(30,5,70,35)
\cut(50,0,50,40)
\end{axopicture}

\\ $F_{17}$: 12 34 013 & $F_{18}$: 12 13 25 124 05 04 & $F_{19}$: 12 13 25 124 013 05 & $F_{20}$: 12 13 124 134 35 25
\\
\num(21)
\Line(10,20)(20,20)
\Line(80,20)(90,20)
\den(20,20,30,35)
\cut(30,35,50,40)
\cut(50,40,70,35)
\cut(70,35,80,20)
\den(70,5,80,20)
\den(50,0,70,5)
\cut(50,0,70,35)
\cut(30,35,50,0)
\dot(60,20)
\den(70,5,50,40)
\dot(56,31)
\dot(35,25)
\den(20,20,70,35)
\end{axopicture}
 & \num(22)
\Line(10,20)(20,20)
\Line(80,20)(90,20)
\den(20,20,30,35)
\cut(30,35,50,40)
\den(50,40,70,35)
\den(70,35,80,20)
\den(20,20,30,5)
\cut(30,5,50,0)
\cut(50,0,70,5)
\cut(70,5,80,20)
\cut(30,5,50,40)
\den(50,0,70,35)
\dot(58,14)
\dot(42,26)
\den(30,35,70,5)
\end{axopicture}
 & \num(23)
\Line(10,20)(20,20)
\Line(80,20)(90,20)
\den(20,20,30,35)
\cut(30,35,50,40)
\den(50,40,70,35)
\den(70,35,80,20)
\den(20,20,30,5)
\den(30,5,50,0)
\cut(50,0,70,5)
\den(70,5,80,20)
\cut(50,0,70,35)
\cut(30,5,50,40)
\cut(30,35,70,5)
\end{axopicture}
 & \num(24)
\Line(10,20)(20,20)
\Line(80,20)(90,20)
\den(20,20,30,35)
\den(30,35,50,40)
\cut(50,40,70,35)
\den(70,35,80,20)
\cut(20,20,30,5)
\den(30,5,50,0)
\cut(50,0,70,5)
\cut(70,5,80,20)
\cut(30,5,50,40)
\den(50,0,70,35)
\dot(42,26)
\dot(58,14)
\den(30,35,70,5)
\end{axopicture}

\\ $F_{21}$: 12 13 24 012 05 & $F_{22}$: 12 13 24 012 45 05 & $F_{23}$: 12 13 25 45 134 012 & $F_{24}$: 12 13 45 134 05 03
\\
\num(25)
\Line(10,20)(20,20)
\Line(80,20)(90,20)
\den(20,20,40,40)
\cut(40,40,60,40)
\den(60,40,80,20)
\cut(20,20,40,0)
\den(40,0,60,0)
\den(60,0,80,20)
\den(40,40,40,20)
\cut(40,20,40,0)
\cut(60,40,60,20)
\cut(60,20,60,0)
\den(40,20,60,20)
\end{axopicture}
 & \num(26)
\Line(10,20)(20,20)
\Line(80,20)(90,20)
\cutd(20,20,50,40)
\den(50,40,80,20)
\den(20,20,50,0)
\cutd(50,0,80,20)

\cut(50,40,50,0)
\end{axopicture}
 & \num(27)
\Line(10,20)(20,20)
\Line(80,20)(90,20)
\den(20,20,40,40)
\den(40,40,60,40)
\cut(60,40,80,20)
\cutd(20,20,40,0)
\den(40,0,60,0)
\den(60,0,80,20)
\cut(40,40,60,0)
\cut(60,40,40,0)
\end{axopicture}
 & \num(28)
\Line(10,20)(20,20)
\Line(80,20)(90,20)
\den(20,20,40,40)
\den(40,40,60,40)
\cut(60,40,80,20)
\cut(20,20,40,0)
\den(40,0,60,0)
\den(60,0,80,20)
\cut(40,40,40,0)
\cut(60,40,60,0)
\cut(60,40,40,0)
\end{axopicture}

\\ $F_{25}$: 12 13 45 134 012 05 & $F_{26}$: 012 123 & $F_{27}$: 12 01 123 013 & $F_{28}$: 123 012 05 01
\\
\num(29)
\Line(10,20)(20,20)
\Line(80,20)(90,20)
\cut(20,20,30,35)
\den(30,35,50,40)
\den(50,40,70,35)
\den(70,35,80,20)
\den(20,20,30,5)
\den(30,5,50,0)
\den(50,0,70,5)
\cut(70,5,80,20)
\cut(30,5,50,40)
\cut(50,0,70,35)
\cut(30,35,70,5)
\end{axopicture}
 & \num(30)
\Line(10,20)(20,20)
\Line(80,20)(90,20)
\den(20,20,30,35)
\cut(30,35,50,40)
\den(50,40,70,35)
\den(70,35,80,20)
\cut(20,20,30,5)
\cut(30,5,50,0)
\den(50,0,70,5)
\cut(70,5,80,20)
\den(30,5,50,40)
\cut(50,0,70,35)
\dot(42,26)
\dot(57,14)
\den(30,35,70,5)
\end{axopicture}
 & \num(31)
\Line(10,20)(20,20)
\Line(80,20)(90,20)
\den(20,20,40,40)
\den(40,40,60,40)
\cut(60,40,80,20)
\cut(20,20,40,0)
\den(40,0,60,0)
\den(60,0,80,20)
\cut(40,40,40,20)
\den(40,20,40,0)
\den(60,40,60,20)
\cut(60,20,60,0)
\cut(40,20,60,20)
\end{axopicture}
 & \num(32)
\end{axopicture}

\\ $F_{29}$: 12 123 25 013 05 01 & $F_{30}$: 12 13 034 134 05 03 & $F_{31}$: 12 13 034 134 05 04 &
\end{tabular}
\caption{Cut diagrams for five-particle phase-space master integrals in QCD.
Dashed lines represent cut propagators and carry final-state momenta
$p_1, \dots, p_5$. Labels represent propagators, so
that "123" corresponds to $p_1+p_2+p_3$ and
"012" to $q-p_1-p_2$ (where $q$ is the initial-state momentum, i.e., $q=p_1+\dots+p_5$).}
\label{tab:mis}
\end{table}

We start by identifying a set of five-particle phase-space master integrals
by constructing auxiliary topologies containing the four-loop massless
propagators from \cite{BC10}, taking all physical five-particle cuts of
those, and performing the Laporta-style IBP reduction \cite{Lap00,Lap17}
implemented in \fire \cite{Smi14}. This gives us 31 master integrals listed in
Table~\ref{tab:mis} with up to 6 unique propagators each. Our notation is
\begin{equation}
    \label{eq:fi}
    F_i = S_\Gamma \int \D \PS_5 ~ \frac{1}{D^{(i)}_1 ~ \dots ~ D^{(i)}_n},
\end{equation}
where $D^{(i)}_j$ are propagators that take the form of invariant scalar
products
\begin{equation}
    s_{kl\dots q} = \left(p_k+p_l+\dots+p_q\right)^2,
\end{equation}
$\D \PS_5$ is a five-particle phase-space element in $D$ dimensions
\begin{equation}
  \label{eq:psint-def}
  \D \PS_N =
  \left (
    \prod_{i=1}^{N} \D^D p_i \x \delta^{+}\big(p_i^2\big)
  \right )
  \delta^{(D)}\Big(q - p_1 - \ldots - p_N \Big),
\end{equation}
and $S_\Gamma$ is a common normalization factor chosen for convenience to be
\begin{equation}
    \label{eq:sgamma}
    S_\Gamma =
        \left(q^2\right)^{5-2D}
        \frac{\big(2\pi\big)^4}{\pi^{2D}}
        \Gamma\left(\frac{D}{2}-1\right)\Gamma\left(3\frac{D}{2}-3\right).
\end{equation}
With this normalization and knowing the volume of the complete $N$-particle
phase space\footnote{The dependence on $q^2$ here is trivial, and can be
restored by power counting. We will omit it from now on, setting $q^2$ to $1$.}
\begin{equation}
    \int \D \PS_N =
        \left(q^2\right)^{\frac{D}{2}(N-1)-N}
        \x
        \frac{\pi^{\frac{D}{2}(N-1)}}{\big(2\pi\big)^{N-1}}
        \frac{\Gamma\left(\frac{D}{2}-1\right)^N}
        {\Gamma\Big(\left(\frac{D}{2}-1\right)\big(N-1\big)\Big)
        \x
        \Gamma\Big(\left(\frac{D}{2}-1\right)N\Big)}
\end{equation}
we can already fix the value of $F_1$ as
\begin{equation}
  \label{eq:f1}
    F_1 = S_\Gamma \int \D \PS_5 =
    \frac{\Gamma\big(\frac{D}{2}-1\big)^6 \x \Gamma\big(3\frac{D}{2}-3\big)}
    {\Gamma\big(4\frac{D}{2}-4\big) \x \Gamma\big(5\frac{D}{2}-5\big)}.
\end{equation}

Next, with the help of \litered \cite{Lee13} and \fire \cite{Smi14} we derive
a set of lowering dimensional recurrence relations which express master
integrals in $D+2$ dimensions in terms of master integrals in $D$ dimensions:
\begin{equation}
    \label{eq:drr}
    F_i(D+2) = M_{ij}(D) \x F_j(D).
\end{equation}
In our case $M$ can be shuffled into triangular form, with each $F_i$ only
depending on itself and master integrals from lower sectors $S_i$:
\begin{equation}
    \label{eq:drr_i}
    F_i(D+2) = M_{ii}(D) \x F_i(D) \x + \sum_{k\in S_i} M_{ik}(D) F_k(D)
\end{equation}
and the general solution being
\begin{equation}
  \label{eq:sol-hom-part}
  F_i(D) = \omega_i(D) \x H_i(D) + R_i(D),
\end{equation}
where $H_i$ is a homogeneous solution of eq.~\eqref{eq:drr_i}, $R_i$ is a
partial solution that can be constructed numerically with \dream~\cite{LM17a}
provided $F_1$ is known, and $ \omega_i(D)$ is an arbitrary periodic
function that needs to be determined from separate considerations.

We argue that all $\omega_{i>1}$ are zero. To see this, first let us look at
the asymptotic behavior of $F_i$ at large $D$. Rewriting eq.~\eqref{eq:fi} as an
integral over invariants $s_{ij}$ gives
\begin{equation}
    \label{eq:fi-inv}
    F_i = S_\Gamma
    \left( \prod_{k=1}^{N-1} \Omega_{D-k} \right)
    \int
        \left (\prod_{l<m} \D s_{lm} \right)
        \big({\Delta_N}\big)^{\frac{D-N-1}{2}}
        \Theta\big(\Delta_N\big)\x
        \delta\left(1 - s_{1\dots N}\right)
        \frac{1}{D^{(i)}_1 ~ \dots ~ D^{(i)}_n},
\end{equation}
where $\Delta_N$ is the Gram determinant defined as
\begin{equation}
  \label{eq:gramdet}
  \Delta_N = \frac{(-1)^{N+1}}{2^N}
  \begin{vmatrix}
    s_{11} & s_{12} & \cdots & s_{1N} \\
    s_{12} & s_{22} & \cdots & s_{2N} \\
    \vdots  & \vdots  & \ddots & \vdots  \\
    s_{1N} & s_{2N} & \cdots & s_{NN}
  \end{vmatrix},
\end{equation}
and $\Omega_{k}$ is the surface of a unit hypersphere in $k$-dimensional space
\begin{equation}
    \Omega_k = 2\pi^{\frac{k}{2}} \x \Gamma\left(\frac{k}{2}\right)^{-1}.
\end{equation}

If $\Delta_N(s_{ij})$ has a unique global maximum inside the integration region, we can
apply Laplace's method to eq.~\eqref{eq:fi-inv} and find its asymptotic as
\begin{equation}
    \label{eq:laplace-asy}
    F_i(D \to \infty) =
        S_\Gamma
        \left( \prod_{k=1}^{N-1} \Omega_{D-k} \right)
        \left( \Delta^{max}_N \right)^{\frac{D}{2}}
        \left( \frac{2\pi}{D} \right)^{\frac{1}{2}\left(\frac{N (N-1)}{2}-1\right)}
        \left( \mathcal{C}_i + \mathcal{O}\left(D^{-1}\right) \right),
\end{equation}
where $\mathcal{C}_i$ is a constant that depends on the location of the maximum
and the denominators $D^{(i)}_j$, but not on $D$.

The global maximum of $\Delta_N$ is reached when all $s_{ij}$ ($i\neq j$) are
identical and equal to $\frac{2}{N (N-1)}$. Geometrically this configuration
corresponds to the vectors $\vec{p}_i$ pointing to the vertices of a regular
$N$-hedron embedded into Euclidean space of $(N-1)$ dimensions. The maximum
value is then
\begin{equation}
    \label{eq:gramdet-max}
    \Delta^{max}_N = \frac{1}{N^{N} (N-1)^{N-1}}
\end{equation}
and explicitly we get
\begin{equation}
    \label{eq:fiasy}
    F_i(D \to \infty) =
        \pi^{\frac{7}{2}} 2^{\frac{25}{2}}
        \frac{\left(4^4 5^5\right)^{-\frac{D}{2}} \Gamma \left(\frac{3 D}{2}-3\right)}
        {D^{\frac{9}{2}}
            \Gamma \left(\frac{D-4}{2}\right)
            \Gamma \left(\frac{D-3}{2}\right)
            \Gamma \left(\frac{D-1}{2}\right)}
        \left( \mathcal{C}_i + \mathcal{O}\left(D^{-1}\right) \right).
\end{equation}
It follows that all $F_i$ have identical asymptotic behavior up
to a constant $\mathcal{C}_i$. As a confirmation, it can be shown that
eq.~\eqref{eq:fiasy} is asymptotically the same expression as we had for $F_1$ in
eq.~\eqref{eq:f1}.

Next, we can find the asymptotics of the homogeneous parts of eq.~\eqref{eq:sol-hom-part},
$H_i(D)$, using e.g. the routine \texttt{FindAsymptotics} from \dream. Comparing
these to eq.~\eqref{eq:fiasy}, we determine that all $H_i(D)$ for $i>1$ are growing
exponentially faster than $F_i(D)$, which can only happen if the corresponding
periodic functions $ \omega_i(D)$ are zero.

Thus, to find $F_i$ we only need to find $R_i$, the inhomogeneous solutions
to eq.~\eqref{eq:drr_i}. We compute them as a series in $\eps=(4-D)/2$ using
\dream with 2000-digit accuracy and then restore the analytical form of the
series coefficients in terms of MZVs using PSLQ method \cite{FBA99}. In this
way we obtain the analytical result for all master integrals up to MZVs of
weight 12 using the corresponding bases from \cite{Fur03} and the \summertime
package \cite{LM15} for their numerical evaluation. Corresponding expressions
are presented in Appendix~\ref{app:res} as well as in the auxiliary files on
arXiv.

\section{Crosschecks}
\label{sec:4}

\subsection{Four-Particle Integrals}

As the first consistency check of our method we reproduce results for four-
particle phase-space integrals reported in \cite{GGH03}. We perform all the
steps described in Section~\ref{sec:2}. Generating the IBP rules with the help
of \litered and then proceeding with \dream we obtain the final result with
2000-digit accuracy and MZVs up to weight 12. The series reconstructed with
PSLQ (using the original notation, and omitting $S_\Gamma$ and $q^2$ factors)
are:
\begin{align}
    R_6 & =  -1+\z(2)
+\eps \Bigg(-12+5 \z(2)+9 \z(3) \Bigg)
+\eps^2 \Bigg(-91+27 \z(2)+45 \z(3)+\frac{61}{5} \z(2)^2\Bigg)
\\ \nn &
+\eps^3 \Bigg(-558
+161 \z(2)+197 \z(3)+61 \z(2)^2-80 \z(3) \z(2)
+207 \z(5)\Bigg)
\\ \nn &
+\eps^4 \Bigg(-3025+939 \z(2)
+897 \z(3)
+\frac{1157}{5} \z(2)^2
-400 \z(3) \z(2) +1035 \z(5)+\frac{288}{5} \z(2)^3-153 \z(3)^2\Bigg)
, \\
    R_{8,a} & = \frac{5}{\eps^4}
-\frac{40 \z(2)}{\eps^2}
-\frac{126 \z(3)}{\eps}
+14 \z(2)^2
+\eps \Bigg(1008 \z(2) \z(3)-1086 \z(5)\Bigg)
+\eps^2 \Bigg(-\frac{272}{7} \z(2)^3+1602 \z(3)^2\Bigg)
, \\
    R_{8,b} & = \frac{3}{4 \eps^4}
-\frac{17 \z(2)}{2 \eps^2}
-\frac{44 \z(3)}{\eps}
-\frac{183}{5} \z(2)^2
+\eps \Bigg(376 \z(2) \z(3)-790 \z(5)\Bigg)
+\eps^2 \Bigg(-\frac{19088}{105} \z(2)^3+698 \z(3)^2\Bigg)
.
\end{align}

\subsection{Numerical Verification}

\begin{table}[h!]
\centering
\begin{tabular}{ rrrrrrr }
 $i$   & \multicolumn{3}{r}{Numerical results}& \multicolumn{3}{r}{Analytic results} \\
\hline
 &      $D=4$ &      $D=6$ &      $D=8$ &      $D=4$ &      $D=6$ &      $D=8$ \\
\hline \\ [-1em]
  2 &         -- &  1708(2)00 &   4699(1)0 &         -- &  171085.62 &  47000.531 \\
  3 &  3.7823(4) &  3.1704(2) &  3.0221(1) &  3.7823736 &  3.1704486 &  3.0221118 \\
  4 &         -- &  1504.7(8) &   725.3(1) &         -- &  1504.4507 &  725.26806 \\
  5 &         -- &  1007.4(5) &  580.80(9) &         -- &  1007.5235 &  580.76347 \\
  6 &         -- &  6191(5)00 &  14496(6)0 &         -- &  619633.25 &  144975.32 \\
  7 &   46.46(4) &  18.533(2) &  15.205(1) &  46.435253 &  18.532303 &  15.205538 \\
  8 &         -- &   2031(2)0 &    5357(2) &         -- &  20297.189 &  5355.3611 \\
  9 &         -- &    4313(3) &  2406.7(4) &         -- &  4312.8823 &  2406.7943 \\
 10 &  10.436(2) &  7.1508(5) &  6.5093(3) &  10.435253 &  7.1507477 &  6.5092878 \\
 11 &   228.8(1) &   62.67(1) &  47.663(4) &  229.11836 &  62.667046 &  47.663194 \\
 12 &         -- &  157.34(4) &  102.26(1) &         -- &  157.33521 &  102.26408 \\
 13 &         -- &   13729(8) &    4000(1) &         -- &  13732.166 &  4000.2779 \\
 14 &         -- &  268.46(8) &  172.80(2) &         -- &  268.45969 &  172.79805 \\
 15 &         -- &   6322(6)0 &   16048(5) &         -- &  63316.356 &  16049.857 \\
 16 &         -- &   4414(3)0 &   12952(4) &         -- &  44117.898 &  12951.443 \\
 17 &         -- &  1243.4(7) &   709.6(1) &         -- &  1243.1369 &  709.52840 \\
 18 &         -- &  3002(2)00 &   5899(3)0 &         -- &  300402.99 &  58965.517 \\
 19 &         -- &  4982(4)00 &  10637(4)0 &         -- &  498329.79 &  106357.81 \\
 20 &         -- & 2360(2)000 &  5777(2)00 &         -- &  2362594.9 &  577686.64 \\
 21 &         -- &   6312(5)0 &   20642(5) &         -- &  63147.876 &  20642.071 \\
 22 &         -- &  8402(7)00 &  24407(8)0 &         -- &  840453.94 &  244075.75 \\
 23 &         -- & 1443(1)000 &  4556(1)00 &         -- &  1443198.3 &  455543.43 \\
 24 &         -- &  1391(1)00 &   3997(1)0 &         -- &  139263.92 &  39966.878 \\
 25 &         -- &  3347(3)00 &   8526(3)0 &         -- &  335128.10 &  85254.217 \\
 26 &  25.563(6) &  15.376(1) & 13.6042(7) &  25.564747 &  15.376404 &  13.604247 \\
 27 &         -- &   697.3(3) &  397.84(6) &         -- &  697.18948 &  397.83514 \\
 28 &   143.9(1) &  52.855(7) &  42.917(3) &  143.97886 &  52.853837 &  42.917424 \\
 29 &         -- &   4409(3)0 &   13702(3) &         -- &  44117.898 &  13700.597 \\
 30 &         -- &   6327(6)0 &   16181(5) &         -- &  63316.356 &  16178.566 \\
 31 &         -- &   8955(8)0 &   19055(8) &         -- &  89611.062 &  19051.115 \\
\hline
\end{tabular}
\caption{%
Numerical results for the ratio $F_i/F_1$ with the corresponding uncertainties
(standard deviations) indicated in the parenthesis. Missing entries correspond
to divergent integrals.}
\label{table:numresults}
\end{table}

As another cross-check we have calculated the leading terms of the
$\epsilon$-expansion of $F_i$ numerically using the direct way: through
Monte-Carlo integration of eq.~\eqref{eq:fi} over the phase-space. While
such a technique can not be easily applied to divergent integrals, we can
sidestep such an issue by noting that our master integrals only suffer from
IR divergences that disappear already at $D=6$. In this way we can check
several leading terms of the expansion at $D=4-2\epsilon$ by calculating the
corresponding integrals in $D\ge6$ since both are connected by dimensional
recurrence relations.

To calculate a finite integral of the form eq.~\eqref{eq:fi} we choose a
uniform mapping from a hypercube into momentum coordinates using an algorithm
similar to RAMBO~\cite{KSE85} but extended into arbitrary $D$. Then we
calculate the integrand from scalar products of the momenta, and finally we
integrate over the hypercube using the Vegas~\cite{Lepage78} implementation
from \textsc{Cuba}~\cite{Hah04}.

Note that although the integrals we are calculating are finite, the integrands
are not. Exposing an integration algorithm like Vegas to such infinities may
lead to unpredictable behaviour, so as a precaution we choose to regulate these
infinities by adding a small parameter $\alpha$ to the denominator of the
integrand, and then to calculate the integral with progressively smaller values
of $\alpha$ (from $2^{-30}$ to $2^{-100}$), checking if convergence was reached
afterwards.

The results of this method are summarized in Table~\ref{table:numresults}, and
show good agreement between numerical and analytic results. Our integration
program is written in C using the GNU Scientific Library~\cite{GSL09} and
\textsc{Cuba}. Its source code can be found at \url{https://hg.tx97.net/rambo},
and also in the auxiliary files on arXiv. With a requested accuracy of $0.1\%$
the complete integration takes less than two days on a 12-core machine, with
each integration taking between a minute and two hours.

\section{Conclusions}
\label{sec:5}
In this paper we present analytical expressions for five-particle phase-space
integrals expressed in terms of multiple zeta values up to weight 12. The
results are calculated using dimensional recurrence relation method with a
2000-digit accuracy using the \dream package. We also present computer code
for the numerical integration of phase-space integrals in a higher-number of
dimensions that has been used to cross-check the obtained results with an
accuracy of $0.1\%$. The approach presented here shows excellent performance
for calculating single-scale integrals without ultra-violet divergences and can
be easily applied to other problems of this kind.

\section{Acknowledgments}
We are thankful to Sven Moch for numerous discussions and helpful suggestions
concerning this work, and for proofreading this paper. We were pleased to use
\texttt{Axodraw2} \cite{CV16} to draw diagrams for this paper.

This work was supported in part by the German Research
Foundation DFG through the Collaborative Research Centre No.\ SFB~676
\textit{Particles, Strings and the Early Universe: the Structure of Matter and
  Space-Time}.

\appendix

\section{Results}
\label{app:res}

The main results of our work are listed below. For brevity, we truncate them up
to MZVs of weight~6. Complete results with weight up to 12 are available in
auxiliary files on arXiv.

\begin{align}
  F_1 & =   \frac{1}{72}
+ \eps\frac{53}{288}
+ \eps^2\Bigg(\frac{15961}{10368}-\frac{13}{72} \z(2)\Bigg)
+ \eps^3\Bigg(
     \frac{436013}{41472}
   - \frac{689}{288} \z(2)
   - \frac{13}{18} \z(3)
  \Bigg)
\\ \nn &
+ \eps^4\Bigg(
     \frac{96102601}{1492992}
   - \frac{207493}{10368} \z(2)
   - \frac{689}{72} \z(3)
   + \frac{17}{240} \z(2)^2
  \Bigg)
+ \eps^5\Bigg(
     \frac{2206279853}{5971968}
   - \frac{5668169}{41472} \z(2)
\\ \nn &
   - \frac{207493}{2592} \z(3)
   + \frac{901}{960} \z(2)^2
   + \frac{169}{18} \z(3) \z(2)
   - \frac{65}{6} \z(5)
\Bigg)
+
\eps^6 \Bigg(
 \frac{437728233961}{214990848}
-\frac{1249333813}{1492992} \z(2)
\\ \nn &
-\frac{5668169}{10368} \z(3)
+\frac{271337}{34560} \z(2)^2
+\frac{8957}{72} \z(3) \z(2)
-\frac{3445}{24} \z(5)
-\frac{4007}{5040} \z(2)^3
+\frac{169}{9} \z(3)^2
\Bigg)

\\
  F_2 & =   \frac{10}{3\eps^5}
- \frac{55}{3\eps^4}
+  \frac{1}{\eps^3}\Bigg(\frac{160}{3}-\frac{130}{3} \z(2)\Bigg)
+ \frac{1}{\eps^2}\Bigg(-\frac{490}{3}+\frac{715}{3} \z(2)-\frac{512}{3} \z(3)\Bigg)
\\ &\nn 
+ \frac{1}{\eps}\Bigg(
     \frac{1450}{3}
   - \frac{2080}{3} \z(2)
   + \frac{2816}{3} \z(3)
   + 25 \z(2)^2
  \Bigg)
  - \frac{4390}{3}
  + \frac{6370}{3} \z(2)
  - \frac{8192}{3} \z(3)
  - \frac{275}{2} \z(2)^2
\\ &\nn 
  + \frac{6656}{3} \z(3) \z(2)
  - 2504 \z(5)
+
\eps \Bigg(
 \frac{13090}{3}
-\frac{18850}{3} \z(2)
+\frac{25088}{3} \z(3)
+400 \z(2)^2
-\frac{36608}{3} \z(3) \z(2)
\\ &\nn 
+13772 \z(5)
-211 \z(2)^3
+\frac{13136}{3} \z(3)^2
\Bigg)

\\
  F_3 & =   \frac{7}{8}-\frac{\z(2)}{2}
+ \eps\Bigg(\frac{219}{16}-\frac{7}{2} \z(2)-6 \z(3)\Bigg)
+ \eps^2\Bigg(\frac{4231}{32}-\frac{227}{8} \z(2)-42 \z(3)-\frac{111}{10} \z(2)^2\Bigg)
\\ &\nn 
+ \eps^3 \Bigg(
     \frac{65347}{64}
   - \frac{3999}{16} \z(2)
   - \frac{499}{2} \z(3)
   - \frac{777}{10} \z(2)^2+86 \z(3) \z(2)-237 \z(5)\Bigg)
+ \eps^4 \Bigg(
     \frac{887695}{128}
\\ &\nn 
   - \frac{64219}{32} \z(2)
   - \frac{6303}{4} \z(3)
   - \frac{5967}{16} \z(2)^2
   + 602 \z(3) \z(2)
   - 1659 \z(5)
   - \frac{1827}{20} \z(2)^3
   + 204 \z(3)^2
  \Bigg)

\\
  F_4 & = -\frac{\z(2)}{\eps^3}
+\frac{1}{\eps^2}\Bigg(\frac{\z(2)}{2}-13 \z(3)\Bigg)
+\frac{1}{\eps}\Bigg(\frac{3}{2} \z(2)+\frac{13}{2} \z(3)-\frac{267}{10} \z(2)^2\Bigg)
+4 \z(2)+\frac{39}{2} \z(3)+\frac{267}{20} \z(2)^2
\\ \nn &
+191 \z(3) \z(2)-\frac{1129}{2} \z(5)
+\eps \Bigg(10 \z(2)+52 \z(3)+\frac{801}{20} \z(2)^2-\frac{191}{2} \z(3) \z(2)+\frac{1129}{4} \z(5)-\frac{16547}{70} \z(2)^3
\\ \nn &
+487 \z(3)^2\Bigg)

\\
  F_5 & =  \frac{7 \z(2)^2}{5\eps^2}
+\frac{1}{\eps}\Bigg( -\frac{7}{2} \z(2)^2-18 \z(3) \z(2)+87 \z(5) \Bigg)
  -\frac{7}{10} \z(2)^2
  +45 \z(3) \z(2)
  -\frac{435}{2} \z(5)
  +\frac{647}{5} \z(2)^3
  -108 \z(3)^2

\\
  F_6 & = -\frac{35}{9\eps^5}
-\frac{19}{6 \eps^4}
+\frac{1}{\eps^3} \Bigg(-\frac{278}{3}-\frac{191}{3} \z(2)\Bigg)
+\frac{1}{\eps^2} \Bigg(\frac{1697}{3}+\frac{1687}{18} \z(2)-\frac{3236}{9} \z(3)\Bigg)
\\ \nn &
+\frac{1}{\eps} \Bigg(-\frac{7793}{3}+\frac{10450}{9} \z(2)+\frac{2386}{3} \z(3) -\frac{4927}{18} \z(2)^2\Bigg)
+\frac{31259}{3}-\frac{67705}{9} \z(2)+4296 \z(3)
\\ \nn &
+\frac{209681}{180} \z(2)^2+\frac{14708}{3} \z(3) \z(2)-\frac{28148}{3} \z(5)
+\eps \Bigg(-\frac{117869}{3}+\frac{317353}{9} \z(2)-31444 \z(3)
\\ \nn &
-\frac{66809}{45} \z(2)^2-\frac{101858}{9} \z(3) \z(2)+\frac{84602}{3} \z(5)-\frac{357871}{126} \z(2)^3+\frac{96472}{9} \z(3)^2\Bigg)

\\
  F_7 & = - 1+\z(2)
+ \eps\Bigg(-17+10 \z(2)+9 \z(3)\Bigg)
+ \eps^2\Bigg(-\frac{351}{2}+\frac{163}{2} \z(2)+90 \z(3)+\frac{36}{5} \z(2)^2\Bigg)
\\ \nn &
+ \eps^3 \Bigg(
    -\frac{5709}{4}
    +\frac{2495}{4} \z(2)
    +\frac{1337}{2} \z(3)
    +72 \z(2)^2
    -151 \z(3) \z(2)
    +207 \z(5)
\Bigg)
+ \eps^4 \Bigg(
    -\frac{80649}{8}
\\ \nn &
    +\frac{35823}{8} \z(2)
    +\frac{18035}{4} \z(3)
    +\frac{4881}{10} \z(2)^2
    -1510 \z(3) \z(2)
    +2070 \z(5)
    -\frac{387}{10} \z(2)^3
    -387 \z(3)^2
\Bigg)

\\
  F_8 & = \frac{1}{2 \x \eps^5} 
-\frac{11}{4 \x\eps^4}
+\frac{1}{\eps^3}\Bigg(8-\frac{49}{6} \z(2)\Bigg)
+\frac{1}{\eps^2}\Bigg(-\frac{49}{2}+\frac{539}{12} \z(2)-\frac{127}{3} \z(3)\Bigg)
+\frac{1}{\eps}\Bigg(\frac{145}{2}-\frac{392}{3} \z(2)
  \\ \nn &
   +\frac{1397}{6} \z(3)-\frac{823}{60} \z(2)^2\Bigg)
  -\frac{439}{2}
  +\frac{2401}{6} \z(2)
  -\frac{2032}{3} \z(3)
  +\frac{9053}{120} \z(2)^2
  +\frac{1829}{3} \z(3) \z(2)
  \\ \nn &
  -\frac{2381}{3} \z(5)
+ \eps \Bigg(
 \frac{1309}{2}
-\frac{7105}{6} \z(2)
+\frac{6223}{3} \z(3)
-\frac{3292}{15} \z(2)^2
-\frac{20119}{6} \z(3) \z(2)
+\frac{26191}{6} \z(5)
  \\ \nn &
+\frac{48983}{1260} \z(2)^3
+1397 \z(3)^2
\Bigg)

\\
  F_9 & = -\frac{\z(2)}{\eps^3}
+\frac{1}{\eps^2}\Bigg(\frac{\z(2)}{2}-7 \z(3)\Bigg)
+\frac{1}{\eps}\Bigg(\frac{3}{2} \z(2)+\frac{7}{2} \z(3)+\frac{17}{10} \z(2)^2\Bigg)
+
 4 \z(2)
+\frac{21}{2} \z(3)
-\frac{17}{20} \z(2)^2
+131 \z(3) \z(2)
\\ \nn &
-\frac{127}{2} \z(5)
+ \eps\Bigg(
 10 \z(2)
+28 \z(3)
-\frac{51}{20} \z(2)^2
-\frac{131}{2} \z(3) \z(2)
+\frac{127}{4} \z(5)
+\frac{7677}{70} \z(2)^3
+328 \z(3)^2
\Bigg)

\\
  F_{10} & = \z(2)-\frac{3}{2}
+ \eps \Bigg(-\frac{47}{2}+\frac{15}{2} \z(2)+11 \z(3)\Bigg)
+ \eps^2 \Bigg(-\frac{903}{4}+58 \z(2)+\frac{165}{2} \z(3) + \frac{171}{10} \z(2)^2\Bigg)
\\ \nn &
+ \eps^3\Bigg(-\frac{13795}{8}
+\frac{951}{2} \z(2)
+\frac{1003}{2} \z(3)
+\frac{513}{4} \z(2)^2
-159 \z(3) \z(2)
+\frac{739}{2} \z(5)
\Bigg)
+ \eps^4 \Bigg(
-\frac{184655}{16}
\\ \nn &
+\frac{14539}{4} \z(2)
+3092 \z(3)
+\frac{6507}{10} \z(2)^2
-\frac{2385}{2} \z(3) \z(2)
+\frac{11085}{4} \z(5)
+\frac{1293}{14} \z(2)^3
-374 \z(3)^2
\Bigg)

\\
  F_{11} & = 10 \z(2) \z(3)-16 \z(5) + \eps\Bigg(-5 \z(3) \z(2)+8 \z(5)-\frac{48}{35} \z(2)^3+43 \z(3)^2\Bigg)

\\
  F_{12} & = -\frac{2 \z(3)}{\eps}
-13 \z(3)-\frac{37}{5} \z(2)^2
+\eps \Bigg(-76 \z(3)-\frac{481}{10} \z(2)^2+30 \z(3) \z(2)-115 \z(5)\Bigg)
+\eps^2 \Bigg(
-422 \z(3)
\\ \nn &
-\frac{1406}{5} \z(2)^2
+195 \z(3) \z(2)
-\frac{1495}{2} \z(5)
-\frac{158}{7} \z(2)^3
+112 \z(3)^2
\Bigg)

\\
  F_{13} & = \frac{1}{6 \eps^5}
-\frac{11}{12 \eps^4}
+\frac{1}{\eps^3}\Bigg(\frac{8}{3}+\frac{\z(2)}{2}\Bigg)
+\frac{1}{\eps^2}\Bigg(-\frac{49}{6}-\frac{11}{4} \z(2)+\frac{86}{3} \z(3)\Bigg)
+\frac{1}{\eps}\Bigg(\frac{145}{6}+8 \z(2)-\frac{473}{3} \z(3)
\\ \nn &
+\frac{5251}{60} \z(2)^2\Bigg)
-\frac{439}{6}
-\frac{49}{2} \z(2)
+\frac{1376}{3} \z(3)
-\frac{57761}{120} \z(2)^2
-466 \z(3) \z(2)
+1794 \z(5)
+\eps\Bigg(
\frac{1309}{6}
\\ \nn &
+\frac{145}{2} \z(2)
-\frac{4214}{3} \z(3)
+\frac{21004}{15} \z(2)^2
+ 2563 \z(3) \z(2)+
-9867 \z(5)
+\frac{1349771}{1260} \z(2)^3
-1372 \z(3)^2
\Bigg)

\\
  F_{14} & = -\frac{8 \z(2)^2}{5 \eps}
-\frac{28}{5} \z(2)^2-48 \z(5)
+ \eps \Bigg(
-\frac{164}{5} \z(2)^2
-168 \z(5)
-\frac{6248}{105} \z(2)^3
+6 \z(3)^2
\Bigg)

\\
  F_{15} & = \frac{1}{6 \eps^6}
+\frac{41}{36\x \eps^5}
+\frac{1}{\eps^4}
\Bigg(
-\frac{311}{36}
-\frac{73}{18} \z(2)
\Bigg)
+\Bigg(\frac{445}{18}-\frac{563}{36} \z(2)-\frac{281}{9} \z(3)\Bigg)\frac{1}{\eps^3}
+\frac{1}{\eps^2}\Bigg(
-\frac{689}{9}
\\ \nn &
+\frac{5273}{36} \z(2)
-\frac{907}{18} \z(3)
-\frac{7103}{180} \z(2)^2
\Bigg)
+\frac{1}{\eps}\Bigg(\frac{2024}{9}-\frac{7759}{18} \z(2)+\frac{13933}{18} \z(3)+\frac{10553}{120} \z(2)^2
\\ \nn &
+\frac{1489}{3} \z(3) \z(2)-\frac{3257}{3} \z(5)\Bigg)
-\frac{6158}{9}
+\frac{12437}{9} \z(2)
-\frac{22193}{9} \z(3)
+\frac{28621}{120} \z(2)^2
+\frac{14065}{18} \z(3) \z(2)
\\ \nn &
+\frac{3631}{6} \z(5)
-\frac{134489}{420} \z(2)^3
+1189 \z(3)^2

\\
  F_{16} & = \frac{1}{6 \eps^6}
+\frac{7}{12 \eps^5}
+\frac{1}{\eps^4}\Bigg(-\frac{185}{36}-\frac{65}{18} \z(2)\Bigg)
+\frac{1}{\eps^3}\Bigg(\frac{209}{18}-\frac{157}{12} \z(2)-\frac{289}{9} \z(3)\Bigg)
+\frac{1}{\eps^2}\Bigg(-\frac{76}{3}
\\ \nn &
+\frac{3563}{36} \z(2)-\frac{215}{2} \z(3)-\frac{ 10927}{180} \z(2)^2\Bigg)
+\frac{1}{\eps}\Bigg(
\frac{239}{9}
-\frac{2123}{18} \z(2)
+\frac{14431}{18} \z(3)
-\frac{ 7041}{40} \z(2)^2
\\ \nn &
+463 \z(3) \z(2)
-1411 \z(5)
\Bigg)
+\frac{857}{9}
-\frac{688}{3} \z(2)
-\frac{5623}{9} \z(3)
+\frac{170591}{120} \z(2)^2
+\frac{8963}{6} \z(3) \z(2)
\\ \nn &
-\frac{8349}{2} \z(5)
-\frac{134141}{180} \z(2)^3
+\frac{4169}{3} \z(3)^2

\\
  F_{17} & = \frac{4 \z(3)}{\eps^2}
+\frac{1}{\eps}\Bigg(10 \z(3)+\frac{84}{5} \z(2)^2\Bigg)
+48 \z(3)+42 \z(2)^2-52 \z(3) \z(2)+280 \z(5)
+\eps \Bigg(
236 \z(3)
+\frac{1008}{5} \z(2)^2
\\ \nn &
-130 \z(3) \z(2)
+700 \z(5)
+\frac{588}{5} \z(2)^3
-160 \z(3)^2
\Bigg)

\\
  F_{18} & = -\frac{5}{12 \eps^6}
+\frac{1}{8 \eps^5}
+\frac{1}{\eps^4}\Bigg(\frac{205}{12}+\frac{323}{36} \z(2)\Bigg)
+\frac{1}{\eps^3}\Bigg(-\frac{1937}{18}-\frac{71}{24} \z(2)+\frac{223}{3} \z(3)\Bigg)
+\frac{1}{\eps^2}\Bigg(\frac{4019}{9}
\\ \nn &
-\frac{3137}{12} \z(2)-\frac{443}{18} \z(3)+\frac{46931}{360} \z(2)^2\Bigg)
+\frac{1}{\eps}\Bigg(-\frac{3047}{2}+\frac{26681}{18} \z(2)-1465 \z(3)-\frac{5473}{144} \z(2)^2
\\ \nn &
-\frac{9113}{9} \z(3) \z(2)
+\frac{27953}{9} \z(5)\Bigg)
+\frac{39460}{9}
-\frac{54149}{9} \z(2)
+\frac{60682}{9} \z(3)
-\frac{33061}{24} \z(2)^2
\\ \nn &
+\frac{6415}{18} \z(3) \z(2)
-\frac{15967}{18} \z(5)
+\frac{12313867}{7560} \z(2)^3
-2302 \z(3)^2

\\
  F_{19} & = -\frac{1}{4 \eps^6}
+\frac{19}{8 \eps^5}
+\frac{1}{\eps^4}\Bigg(\frac{64}{9}+\frac{173}{36} \z(2)\Bigg)
+\frac{1}{\eps^3}\Bigg(-\frac{1895}{18}-\frac{ 2767}{72} \z(2)+\frac{371}{9} \z(3)\Bigg)
+\frac{1}{\eps^2}\Bigg(
\frac{9157}{18}
\\ \nn &
-\frac{1451}{18} \z(2)
-\frac{3439}{18} \z(3)
+\frac{32357}{360} \z(2)^2
\Bigg)
+\frac{1}{\eps}\Bigg(
-\frac{5798}{3}
+\frac{24635}{18} \z(2)
-\frac{7103}{18} \z(3)
-\frac{17023}{720}  \z(2)^2
\\ \nn &
-\frac{4127}{9} \z(3) \z(2)
+\frac{18073}{9} \z(5)
\Bigg)
+\frac{111845}{18}
-\frac{120679}{18} \z(2)
+\frac{49330}{9} \z(3)
-\frac{3664}{5} \z(2)^2
+\frac{50743}{18} \z(3) \z(2)
\\ \nn &
-\frac{53357}{18} \z(5)
+\frac{10750309}{7560} \z(2)^3
-\frac{7078}{9} \z(3)^2

\\
  F_{20} & = \frac{20}{9 \eps^6}
+\frac{85}{6 \eps^5}
+\frac{1}{\eps^4}\Bigg(-\frac{775}{12}-\frac{325}{9} \z(2)\Bigg)
+\frac{1}{\eps^3}\Bigg(-\frac{985}{18}-\frac{1705}{9} \z(2)-\frac{1595}{9} \z(3)\Bigg)
\\ \nn &
+\frac{1}{\eps^2}\Bigg(\frac{6445}{9}
-\frac{88}{9} \z(2)^2+\frac{3815}{4} \z(2)-\frac{4775}{6} \z(3)\Bigg)
+\frac{1}{\eps}\Bigg(-3635+\frac{8935}{18} \z(2)+\frac{26455}{6} \z(3)
\\ \nn &
-\frac{743}{36} \z(2)^2
+\frac{23485}{9} \z(3) \z(2)-\frac{7535}{3} \z(5)\Bigg)
+\frac{121310}{9}-\frac{80005}{9} \z(2)+\frac{7415}{9} \z(3)+\frac{9299}{24} \z(2)^2
\\ \nn &
+\frac{189155}{18} \z(3) \z(2)
-\frac{26065}{2} \z(5)+\frac{134581}{126} \z(2)^3+\frac{54985}{9} \z(3)^2

\\
  F_{21} & = \frac{1}{2 \eps^5}
-\frac{11}{4 \eps^4}
+\frac{1}{\eps^3}\Bigg(8+\frac{3}{2} \z(2)\Bigg)
+\frac{1}{\eps^2}\Bigg(-\frac{49}{2}-\frac{ 33}{4}\z(2)+\frac{158}{3} \z(3)\Bigg)
+\frac{1}{\eps}\Bigg(
 \frac{145}{2}
+24 \z(2)
\\ \nn &
-\frac{869}{3} \z(3)
+\frac{1205}{12} \z(2)^2
\Bigg)
-\frac{439}{2}
-\frac{147}{2} \z(2)
+\frac{2528}{3} \z(3)
-\frac{ 13255}{24} \z(2)^2
-\frac{2326}{3} \z(3) \z(2)
+\frac{5614}{3} \z(5)
\\ \nn &
+\eps \Bigg(
 \frac{1309}{2}
+\frac{435}{2} \z(2)
-\frac{7742}{3} \z(3)
+\frac{4820}{3} \z(2)^2
+\frac{12793}{3} \z(3) \z(2)
-\frac{30877}{3} \z(5)
+\frac{565729}{1260} \z(2)^3
\\ \nn &
-\frac{5108}{3} \z(3)^2
\Bigg)

\\
  F_{22} & = \frac{23}{9 \eps^6}
+\frac{55}{9 \eps^5}
+\frac{1}{\eps^4}\Bigg(-\frac{2521}{36}-\frac{127}{3} \z(2)\Bigg)
+\frac{1}{\eps^3}\Bigg(\frac{1777}{9}-\frac{190}{3} \z(2)-\frac{692}{3} \z(3)\Bigg)
\\ \nn &
+\frac{1}{\eps^2}\Bigg(-\frac{11117}{18}
+\frac{11419}{12} \z(2)-\frac{400 }{3}\z(3)-\frac{11507}{90} \z(2)^2\Bigg)
+\frac{1}{\eps}\Bigg(
 \frac{32441}{18}
-\frac{7993}{3} \z(2)
\\ \nn &
+\frac{12061}{3} \z(3)
+\frac{6109}{18} \z(2)^2
+\frac{28472}{9} \z(3) \z(2)
-\frac{15326}{3} \z(5)
\Bigg)
-\frac{99143}{18}
+\frac{50243}{6} \z(2)
-\frac{33388}{3} \z(3)
\\ \nn &
-\frac{16091}{360} \z(2)^2
+\frac{13690}{9} \z(3) \z(2)
+\frac{5165}{3} \z(5)
-\frac{ 539759}{630} \z(2)^3
+\frac{61504}{9} \z(3)^2

\\
  F_{23} & = \frac{115}{18 \eps^6}
-\frac{115}{36 \eps^5}
+\frac{1}{\eps^4}\Bigg(-\frac{2645}{36}-\frac{ 1345}{18} \z(2)\Bigg)
+\frac{1}{\eps^3}\Bigg(\frac{3565}{18}+\frac{1345}{36} \z(2)-\frac{2255}{9} \z(3)\Bigg)
\\ \nn &
+\frac{1}{\eps^2}\Bigg(
-\frac{5750}{9}
+\frac{30935}{36} \z(2)
+\frac{2255}{18} \z(3)
+\frac{1235}{12} \z(2)^2
\Bigg)
+\frac{1}{\eps}\Bigg(
 \frac{16445}{9}
-\frac{41695}{18} \z(2)
\\ \nn &
+\frac{51865}{18} \z(3)
-\frac{ 1235}{24} \z(2)^2
+\frac{26705}{9} \z(3) \z(2)
-\frac{9910}{3} \z(5)
\Bigg)
-\frac{50945}{9}
+\frac{67250}{9} \z(2)
-\frac{69905}{9} \z(3)
\\ \nn &
-\frac{28405}{24} \z(2)^2
-\frac{26705}{18} \z(3) \z(2)
+\frac{4955}{3} \z(5)
-\frac{280151}{252} \z(2)^3
+\frac{45385}{9} \z(3)^2

\\
  F_{24} & = \frac{1}{3 \eps^6}
+\frac{17}{9 \eps^5}
+\frac{1}{\eps^4}\Bigg(-\frac{179}{12}-\frac{67}{9} \z(2)\Bigg)
+\frac{1}{\eps^3}\Bigg(\frac{370}{9}-\frac{89}{3} \z(2)-\frac{568}{9} \z(3)\Bigg)
+\frac{1}{\eps^2}\Bigg(
-\frac{2197}{18}
\\ \nn &
+\frac{9253}{36} \z(2)
-\frac{466}{3} \z(3)
-\frac{ 9659}{90} \z(2)^2
\Bigg)
+\frac{1}{\eps}\Bigg(
 \frac{669}{2}
-\frac{5840}{9} \z(2)
+\frac{14671}{9} \z(3)
-\frac{ 6923}{90} \z(2)^2
\\ \nn &
+\frac{2836}{3} \z(3) \z(2)
-2610 \z(5)
\Bigg)
-\frac{16739}{18}
+\frac{3441}{2} \z(2)
-\frac{33956}{9} \z(3)
+\frac{628781}{360} \z(2)^2
+\frac{18284}{9} \z(3) \z(2)
\\ \nn &
-3109 \z(5)
-\frac{ 746701}{630} \z(2)^3
+\frac{7664}{3} \z(3)^2

\\
  F_{25} & = \frac{14}{9 \eps^5}
-\frac{1}{\eps^4}
+\frac{1}{\eps^3}\Bigg(-\frac{358}{9}-\frac{203}{9} \z(2)\Bigg)
+\frac{1}{\eps^2}\Bigg(
 \frac{2182}{9}
+\frac{187}{6} \z(2)
-\frac{1159}{9} \z(3)
\Bigg)
\\ \nn &
+\frac{1}{\eps}\Bigg(
-\frac{10072}{9}
+\frac{3946}{9} \z(2)
+\frac{5333}{18} \z(3)
-\frac{ 6451}{45} \z(2)^2
\Bigg)
+\frac{40714}{9}
-\frac{25759}{9} \z(2)
+1442 \z(3)
\\ \nn &
+\frac{48637}{90} \z(2)^2
+\frac{13613}{9} \z(3) \z(2)
-3923 \z(5)
+\eps \Bigg(
-\frac{154768}{9}
+\frac{122299}{9} \z(2)
-10793 \z(3)
\\ \nn &
-\frac{19313}{45} \z(2)^2
-\frac{21757}{6} \z(3) \z(2)
+\frac{22801}{2} \z(5)
-\frac{3443939}{1890} \z(2)^3
+\frac{25937}{9} \z(3)^2
\Bigg)

\\
  F_{26} & =  2-\z(2)
+\eps \Bigg(32-9 \z(2)-10 \z(3)\Bigg)
+\eps^2 \Bigg(313
-\frac{165}{2} \z(2)-90 \z(3)
-\frac{ 111}{10} \z(2)^2
\Bigg)
+\eps^3 \Bigg(
\frac{4855}{2}
\\ \nn &
-\frac{2905}{4} \z(2)
-669 \z(3)
-\frac{999}{10} \z(2)^2
+164 \z(3) \z(2)
-268 \z(5)
\Bigg)
+\eps^4 \Bigg(\frac{65815}{4}
-\frac{45421}{8} \z(2)
\\ \nn &
-\frac{9533}{2} \z(3)
-\frac{12339}{20} \z(2)^2
+1476 \z(3) \z(2)
-2412 \z(5)
+\frac{1383}{70} \z(2)^3
+430 \z(3)^2
\Bigg)

\\
  F_{27} & = \frac{2 \z(3)}{\eps^2}
+\frac{1}{\eps}\Bigg(-\z(3)+\frac{52}{5} \z(2)^2\Bigg)
-3 \z(3)-\frac{26}{5} \z(2)^2-8 \z(3) \z(2)+178 \z(5)
+\eps \Bigg(
-8 \z(3)
-\frac{78}{5} \z(2)^2
\\ \nn &
+4 \z(3) \z(2)
-89 \z(5)
+\frac{5342}{35} \z(2)^3
+10 \z(3)^2
\Bigg)

\\
  F_{28} & = -\frac{\z(2)^3}{5}+2 \z(3)^2

\\
  F_{29} & = \frac{5}{9 \eps^6}
-\frac{25}{18 \eps^5}
+\frac{1}{\eps^4}\Bigg(-\frac{5}{18}-\frac{55}{9} \z(2)\Bigg)
+\frac{1}{\eps^3}\Bigg(
-\frac{5}{9}
+\frac{371}{18} \z(2)
-\frac{10}{9} \z(3)
\Bigg)
+\frac{1}{\eps^2}\Bigg(-\frac{10}{9}
\\ \nn &
-\frac{425}{18} \z(2)
+\frac{949}{9} \z(3)+\frac{9167}{90} \z(2)^2\Bigg)
+\frac{1}{\eps}\Bigg(
-\frac{20}{9}
+\frac{595}{9} \z(2)
-\frac{4597}{9} \z(3)
+\frac{11021}{180} \z(2)^2
+\frac{578}{9} \z(3) \z(2)
\\ \nn &
+\frac{4357}{3} \z(5)
\Bigg)
-\frac{40}{9}
-\frac{946}{9} \z(2)
+\frac{10234}{9} \z(3)
-\frac{292403}{180} \z(2)^2
-\frac{10673}{9} \z(3) \z(2)
+\frac{4225}{2} \z(5)
\\ \nn &
+\frac{797803}{630} \z(2)^3
+\frac{430}{9} \z(3)^2

\\
  F_{30} & = \frac{1}{6 \eps^6}
+\frac{13}{12 \eps^5}
+\frac{1}{\eps^4}\Bigg(-\frac{25}{3}-\frac{25}{6} \z(2)\Bigg)
+\frac{1}{\eps^3}\Bigg(\frac{143}{6}-\frac{181}{12} \z(2)-\frac{98}{3} \z(3)\Bigg)
+\frac{1}{\eps^2}\Bigg(-\frac{443}{6}
\\ \nn &
+\frac{431}{3} \z(2)-\frac{169}{3} \z(3)-\frac{163}{4} \z(2)^2\Bigg)
+\frac{1}{\eps}\Bigg(
 \frac{1301}{6}
-\frac{833}{2} \z(2)
+802 \z(3)
+\frac{5887}{120} \z(2)^2
+\frac{1606}{3} \z(3) \z(2)
\\ \nn &
-\frac{3344}{3} \z(5)
\Bigg)
-\frac{3959}{6}
+\frac{7979}{6} \z(2)
-\frac{7286}{3} \z(3)
+\frac{5891}{15} \z(2)^2
+797 \z(3) \z(2)
+\frac{8}{3} \z(5)
-\frac{348377}{1260} \z(2)^3
\\ \nn &
+\frac{4240}{3} \z(3)^2

\\
  F_{31} & =  \frac{7}{9 \eps^5}
-\frac{17}{18 \eps^4}
+\frac{1}{\eps^3}\Bigg(-\frac{143}{9}-\frac{ 125}{9} \z(2)\Bigg)
+\frac{1}{\eps^2}\Bigg(
 \frac{902}{9}
+\frac{133}{6} \z(2)
-\frac{236}{3} \z(3)
\Bigg)
+\frac{1}{\eps}\Bigg(-\frac{4190}{9}
\\ \nn &
+\frac{716}{3} \z(2)
+\frac{1418}{9} \z(3)
-\frac{265}{6} \z(2)^2
\Bigg)
+\frac{16892}{9}
-\frac{4709}{3} \z(2)
+\frac{9718}{9} \z(3)
+\frac{3373}{20} \z(2)^2
+1228 \z(3) \z(2)
\\ \nn &
-\frac{17612}{9} \z(5)
+\eps \Bigg(
-\frac{63902}{9}
+\frac{22181}{3} \z(2)
-\frac{68062}{9} \z(3)
-\frac{377}{5} \z(2)^2
-\frac{23666}{9} \z(3) \z(2)
+\frac{48610}{9} \z(5)
\\ \nn &
-\frac{688249}{1890} \z(2)^3
+\frac{27128}{9} \z(3)^2
\Bigg)

\end{align}


\begin{thebibliography}{AMV12}

\bibitem[AMV12]{AMV11}
A.~A. Almasy, S.~Moch, and A.~Vogt.
\newblock {On the Next-to-Next-to-Leading Order Evolution of Flavour-Singlet
  Fragmentation Functions}.
\newblock {\em Nucl. Phys.}, B854:133--152, 2012.
\newblock \href {http://arxiv.org/abs/1107.2263} {\path{arXiv:1107.2263}},
  \href {http://dx.doi.org/10.1016/j.nuclphysb.2011.08.028}
  {\path{doi:10.1016/j.nuclphysb.2011.08.028}}.

\bibitem[BBV10]{BBV09}
J.~Blumlein, D.~J. Broadhurst, and J.~A.~M. Vermaseren.
\newblock {The Multiple Zeta Value Data Mine}.
\newblock {\em Comput. Phys. Commun.}, 181:582--625, 2010.
\newblock \href {http://arxiv.org/abs/0907.2557} {\path{arXiv:0907.2557}},
  \href {http://dx.doi.org/10.1016/j.cpc.2009.11.007}
  {\path{doi:10.1016/j.cpc.2009.11.007}}.

\bibitem[BC10]{BC10}
P.~A. Baikov and K.~G. Chetyrkin.
\newblock {Four Loop Massless Propagators: An Algebraic Evaluation of All
  Master Integrals}.
\newblock {\em Nucl. Phys.}, B837:186--220, 2010.
\newblock \href {http://arxiv.org/abs/1004.1153} {\path{arXiv:1004.1153}},
  \href {http://dx.doi.org/10.1016/j.nuclphysb.2010.05.004}
  {\path{doi:10.1016/j.nuclphysb.2010.05.004}}.

\bibitem[CT81]{CT81}
K.~G. Chetyrkin and F.~V. Tkachov.
\newblock {Integration by Parts: The Algorithm to Calculate beta Functions in 4
  Loops}.
\newblock {\em Nucl. Phys.}, B192:159--204, 1981.
\newblock \href {http://dx.doi.org/10.1016/0550-3213(81)90199-1}
  {\path{doi:10.1016/0550-3213(81)90199-1}}.

\bibitem[CV16]{CV16}
J.~C. Collins and J.~A.~M. Vermaseren.
\newblock {Axodraw Version 2}.
\newblock 2016.
\newblock \href {http://arxiv.org/abs/1606.01177} {\path{arXiv:1606.01177}}.

\bibitem[FBA99]{FBA99}
H.~Ferguson, D.~Bailey, and S.~Arno.
\newblock Analysis of {PSLQ}, an integer relation finding algorithm.
\newblock {\em Mathematics of Computation of the American Mathematical
  Society}, 68(225):351--369, 1999.
\newblock \href {http://dx.doi.org/10.1090/S0025-5718-99-00995-3}
  {\path{doi:10.1090/S0025-5718-99-00995-3}}.

\bibitem[Fur03]{Fur03}
H.~Furusho.
\newblock The multiple zeta value algebra and the stable derivation algebra.
\newblock {\em Publications of the Research Institute for Mathematical
  Sciences}, 39(4):695--720, 2003.
\newblock \href {http://arxiv.org/abs/math/0011261}
  {\path{arXiv:math/0011261}}, \href
  {http://dx.doi.org/10.2977/prims/1145476044}
  {\path{doi:10.2977/prims/1145476044}}.

\bibitem[GGH04]{GGH03}
A.~{Gehrmann-De~Ridder}, T.~Gehrmann, and G.~Heinrich.
\newblock {Four particle phase space integrals in massless QCD}.
\newblock {\em Nucl.Phys.}, B682:265--288, 2004.
\newblock \href {http://arxiv.org/abs/hep-ph/0311276}
  {\path{arXiv:hep-ph/0311276}}, \href
  {http://dx.doi.org/10.1016/j.nuclphysb.2004.01.023}
  {\path{doi:10.1016/j.nuclphysb.2004.01.023}}.

\bibitem[Git16]{Git15}
O.~Gituliar.
\newblock {Master integrals for splitting functions from differential equations
  in QCD}.
\newblock {\em JHEP}, 02:017, 2016.
\newblock \href {http://arxiv.org/abs/1512.02045} {\path{arXiv:1512.02045}},
  \href {http://dx.doi.org/10.1007/JHEP02(2016)017}
  {\path{doi:10.1007/JHEP02(2016)017}}.

\bibitem[GM15]{GM15}
O.~Gituliar and S.~Moch.
\newblock {Towards three-loop QCD corrections to the time-like splitting
  functions}.
\newblock {\em Acta Phys. Polon.}, B46(7):1279--1289, 2015.
\newblock \href {http://arxiv.org/abs/1505.02901} {\path{arXiv:1505.02901}},
  \href {http://dx.doi.org/10.5506/APhysPolB.46.1279}
  {\path{doi:10.5506/APhysPolB.46.1279}}.

\bibitem[Gou09]{GSL09}
B.~Gough.
\newblock {\em GNU Scientific Library Reference Manual}.
\newblock Network Theory Ltd., 3rd edition, 2009.

\bibitem[Hah05]{Hah04}
T.~Hahn.
\newblock {CUBA: A Library for multidimensional numerical integration}.
\newblock {\em Comput. Phys. Commun.}, 168:78--95, 2005.
\newblock \href {http://arxiv.org/abs/hep-ph/0404043}
  {\path{arXiv:hep-ph/0404043}}, \href
  {http://dx.doi.org/10.1016/j.cpc.2005.01.010}
  {\path{doi:10.1016/j.cpc.2005.01.010}}.

\bibitem[KSE86]{KSE85}
R.~Kleiss, W.~J. Stirling, and S.~D. Ellis.
\newblock {A New Monte Carlo Treatment of Multiparticle Phase Space at
  High-energies}.
\newblock {\em Comput. Phys. Commun.}, 40:359, 1986.
\newblock \href {http://dx.doi.org/10.1016/0010-4655(86)90119-0}
  {\path{doi:10.1016/0010-4655(86)90119-0}}.

\bibitem[Lap00]{Lap00}
S.~Laporta.
\newblock {High precision calculation of multiloop Feynman integrals by
  difference equations}.
\newblock {\em Int. J. Mod. Phys.}, A15:5087--5159, 2000.
\newblock \href {http://arxiv.org/abs/hep-ph/0102033}
  {\path{arXiv:hep-ph/0102033}}, \href
  {http://dx.doi.org/10.1016/S0217-751X(00)00215-7, 10.1142/S0217751X00002157}
  {\path{doi:10.1016/S0217-751X(00)00215-7, 10.1142/S0217751X00002157}}.

\bibitem[Lap17]{Lap17}
S.~Laporta.
\newblock {High-precision calculation of the 4-loop contribution to the
  electron g-2 in QED}.
\newblock {\em Phys. Lett.}, B772:232--238, 2017.
\newblock \href {http://arxiv.org/abs/1704.06996} {\path{arXiv:1704.06996}},
  \href {http://dx.doi.org/10.1016/j.physletb.2017.06.056}
  {\path{doi:10.1016/j.physletb.2017.06.056}}.

\bibitem[Lee14]{Lee13}
R.~N. Lee.
\newblock {LiteRed 1.4: a powerful tool for reduction of multiloop integrals}.
\newblock {\em J. Phys. Conf. Ser.}, 523:012059, 2014.
\newblock \href {http://arxiv.org/abs/1310.1145} {\path{arXiv:1310.1145}},
  \href {http://dx.doi.org/10.1088/1742-6596/523/1/012059}
  {\path{doi:10.1088/1742-6596/523/1/012059}}.

\bibitem[Lep78]{Lepage78}
G.~P. Lepage.
\newblock {A New Algorithm for Adaptive Multidimensional Integration}.
\newblock {\em J. Comput. Phys.}, 27:192, 1978.
\newblock \href {http://dx.doi.org/10.1016/0021-9991(78)90004-9}
  {\path{doi:10.1016/0021-9991(78)90004-9}}.

\bibitem[LM16]{LM15}
R.~N. Lee and K.~T. Mingulov.
\newblock {Introducing SummerTime: a package for high-precision computation of
  sums appearing in DRA method}.
\newblock {\em Comput. Phys. Commun.}, 203:255--267, 2016.
\newblock \href {http://arxiv.org/abs/1507.04256} {\path{arXiv:1507.04256}},
  \href {http://dx.doi.org/10.1016/j.cpc.2016.02.018}
  {\path{doi:10.1016/j.cpc.2016.02.018}}.

\bibitem[LM17a]{LM17a}
R.~N. Lee and K.~T. Mingulov.
\newblock {DREAM, a program for arbitrary-precision computation of dimensional
  recurrence relations solutions, and its applications}.
\newblock 2017.
\newblock \href {http://arxiv.org/abs/1712.05173} {\path{arXiv:1712.05173}}.

\bibitem[LM17b]{LM17b}
R.~N. Lee and K.~T. Mingulov.
\newblock {Meromorphic solutions of recurrence relations and DRA method for
  multicomponent master integrals}.
\newblock 2017.
\newblock \href {http://arxiv.org/abs/1712.05166} {\path{arXiv:1712.05166}}.

\bibitem[Smi15]{Smi14}
A.~V. Smirnov.
\newblock {FIRE5: a C++ implementation of Feynman Integral REduction}.
\newblock {\em Comput. Phys. Commun.}, 189:182--191, 2015.
\newblock \href {http://arxiv.org/abs/1408.2372} {\path{arXiv:1408.2372}},
  \href {http://dx.doi.org/10.1016/j.cpc.2014.11.024}
  {\path{doi:10.1016/j.cpc.2014.11.024}}.

\bibitem[Tar96]{Tar96}
O.~V. Tarasov.
\newblock {Connection between Feynman integrals having different values of the
  space-time dimension}.
\newblock {\em Phys. Rev.}, D54:6479--6490, 1996.
\newblock \href {http://arxiv.org/abs/hep-th/9606018}
  {\path{arXiv:hep-th/9606018}}, \href
  {http://dx.doi.org/10.1103/PhysRevD.54.6479}
  {\path{doi:10.1103/PhysRevD.54.6479}}.

\end{thebibliography}

\end{document}